\documentclass[referee]{raa}          

\usepackage{graphicx,times}             
\input{epsf.sty}                        
\input{psfig.sty}                       

\begin{document}

\title{Variable Stars in the Field of the Open Cluster NGC 2126}
  \volnopage{Vol.0 (200x) No.0, 000--000}
  \setcounter{page}{1}

  \author{S.-F. Liu
     \inst{1,2}
   \and Z.-Y. Wu
     \inst{1}
   \and X.-B. Zhang
     \inst{1}
   \and J.-H. Wu
     \inst{1}
   \and J. Ma
      \inst{1}
    \and Z.-J. Jiang
       \inst{1}
     \and J.-S. Chen
        \inst{1}
     \and X. Zhou
        \inst{1}
        }

  \institute{National Astronomical Observatories, Chinese Academy of
Sciences, Beijing 100012, China; {\it zhouxu@bao.ac.cn}\\
    \and
        Graduate University of Chinese Academy of Sciences,
Beijing 100049, China\\
            }
\date{Received~~2009 month day; accepted~~2009~~month day}

\abstract{We report the results of a time-series CCD photometric
survey of variable stars in the field of the open cluster NGC 2126.
In about one square degree field covering the cluster, a total number
of 21 variable candidates are detected during this survey, of which
16 are newly found. The periods, classifications and spectral types
of 14 newly discovered variables are discussed, which consist of six
eclipsing binaries systems, three pulsating variable stars, three
long period variables, one RS CVn star, and one W UMa or
$\delta$ Scuti star. In addition, there are two variable candidates, the
properties of which cannot be determined in this paper. By a method
based on fitting spectral energy distributions(SEDs) of stars with
theoretical ones, the membership probabilities and the fundamental
parameters of this cluster are determined. As a result, five variables are probably members of NGC 2126. The fundamental parameters of the this cluster
are determined as: the metallicity to be 0.008 $Z_\odot$, the
age $\log(t)$= 8.95, the distance modulus $(m-M)_{0}$= 10.34 and the
reddening value $E(B-V)$=0.55 mag. 
   }
   \authorrunning{S.-F. Liu, Z.-Y. Wu, X.-B. Zhang \& X. Zhou }
   \titlerunning{ Variable Stars in the Field of the Open Cluster NGC 2126 (I)}

   \maketitle
\section{Introduction}

\label{sect:intro} 

Studies of variable stars in star clusters are crucial
tools for understanding of stellar evolution and the nature of the
host clusters. Recent works on variable stars in clusters can be
found in Frandsen \& Arentoft~(\cite{fran98}); Park \&
Nemec~(\cite{par00}); Zloczewski et al.~(\cite{zlo07}). Population I
pulsating variables such as delta Scuti stars, $\gamma$ Dor stars
and slow pulsating B stars are populous in open clusters, and
Population II pulsating variables such as RR Lyr-type stars and SX
Phoenicis-type stars are rich in globular clusters. Stars in open
clusters are not as crowed as globular clusters, so high-precision
photometric results for open clusters can be obtained. Therefore,
open clusters are very important targets to investigate pulsating
variables. Eclipsing binary stars are also found in open clusters.
For examples, 45 short-period eclipsing binaries were discovered in
open cluster Collinder 61 (Mazur et al ~\cite{maz95}), seven
eclipsing binaries are detected in the open cluster NGC 2099 (Kang
et al.~\cite{kang07}), sixty-one eclipsing systems were detected in
the open cluster NGC 6791 (Marchi et al.~\cite{mar07}).

The Beijing-Arizona-Taiwan-Connecticut (BATC) photometric system is
a multicolor sky survey. The system has shown the advantages for open
cluster studies (Fan et al.~\cite{fan96}; Wu et
al.~\cite{wu05},~\cite{wu07}). This system has also contributes to
search for variable stars in clusters. For instances, seventeen
pulsating stars and ten eclipsing systems were found in open
clusters NGC 188 (Zhang et al.~\cite{zhang02},~\cite{zhang04}).
Three pulsating variable stars were discovered in the direction
toward NGC 4565 (Li et al.~\cite{li04}). Six eclipsing binaries and
seven pulsating variables in the field of open cluster 2168 (Hu et
al.~\cite{hu05}). In this paper, we search for variable stars in the
field of NGC 2126.

NGC 2126 ($\alpha_{2000}$=$06^h 02^m$.55, $\delta_{2000}$=
+49$^\circ$52$^\prime$; $l$=163.23885, $b$=13.13066$^\circ$ ) is a
moderately rich typical open cluster with several dozens of
memberships scattered in a region of 5$^\prime$ $\sim$ 6$^\prime$ in
the constellation Auriga (Lyng{\aa}~\cite{lyn87}). The first
photometry was made in blue and red ($\lambda$$\lambda$ 4300-6200
\AA) photographic observation by Cuffey~(\cite{cuf43}), who
estimated the distance to NGC 2126 as 950 pc. Sixty years
later, the first CCD $V R_c$ $I_c$ photometric observation was
performed by G\'asp\'ar et al.~(\cite{a.g03}). They found 6 variable
stars in the cluster field, of which three could be members. They
also obtained the fundamental parameters of the cluster: the age is
9.0$<$$\log(t)$$<$9.4, the reddening value $E(B-V)=0^{m}.2\pm0.^m15$, and the
distance $d\leq 1.3\pm0.6$ kpc. To search for more cluster variable
stars in the open cluster NGC 2126, we performed a dedicated CCD
survey with the BATC system. In this paper, observations and
data reduction are described in Section 2. Searching for
variable stars is reported in Section 3. The physical parameters and the CMD of the Cluster are described in Section 4. Results and discussions are
presented in Section 5. A brief summary is given in Section 6.

\section{Observations and Data Reduction}
\subsection{Observations}
We observed NGC 2126 from 2004 December 4 to 2005 March 24. The
observations were conducted with the 60/90 cm f/3 Schmidt telescope,
which located at the Xinglong Station of the National Astronomical
Observatories, Chinese Academy of Science (NAOC), with a
2K$\times$2K CCD camera at its main focus. The field view of the CCD
covers about 1 \time 1 deg$^{2}$, with an image scale of
1.7$\arcsec$ per pixel. The filter system consists of 15
intermediate-band filters with bandwidths 350 \AA-490 \AA, covering
a spectral range 3000-10000\AA, which are specifically designed to
avoid most of the known bright and variable night-sky emission
lines. In searching for variable stars observations, the {\it f} (5270\AA ) and {\it i}
(6660 \AA) band filters were employed. The
exposure times were set as 300 seconds in {\it f}-band and 120 seconds in the {\it i}-band
in each observation. In total, we got 487
frames of NGC 2126, 243 in {\it f}-band and 244 in {\it i}-band,
respectively.

\subsection{Data Reduction}

Reduction of the CCD data including bias substraction and
flat-fielding with dome flats was performed with an automatic
procedure called PIPELINE I, which has been developed as a standard
programme for BATC multicolor Sky Survey (Fan et al.~\cite{fan96}).
The dome flat-field images were taken by using a diffuser plate in
front of the correcting plate of the Schmidt telescope, the flat
fielding technique has been verified (Wu et al.~\cite{wu02}; Zhou et
al.~\cite{zhou04}). The instrumental magnitudes of point sources in
CCD frames were measured by a PIPELINE II program, which is based on
the DAOPHOT stellar photometric reduction package of
Stetson~(\cite{stet87}). The PIPELINE II program was performed on
each single CCD frame to get magnitude of each point source by PSF
(point spread function) fitting and the aperture photometries with
different aperture. As a result, the list of PSF and aperture
magnitudes in various filter bands were obtained for all the objects
detected.

\subsection{Flux Calibration}

In order to give the spectral types of the variables (see \S 3.1),
we observed NGC 2126 with another 13 BATC filter bands (from $c$ to
$p$) in one night of semi-photometric conditions. Due to not being
enough nights of photometric conditions, the flux calibrations of
NGC 2126 are only defined in {\it f}-band and {\it i}-band by the
four standard stars (Oke \& Gunn~\cite{oke83}) (BD+174708,
BD+262606, HD19445 and HD849373). The flux of the four stars have
been recalibrated by Fukugita et al.~(\cite{fuk96}). In this paper,
the standard stars were observed between air masses of 1.0 and 2.0.
The details of the flux calibration were described in Zhou et
al.~(\cite{zhou01}). Since the flux calibration of NGC 2126 in {\it
f}-band and {\it i}-band have been defined by the four standard
stars as described above, we can perform flux calibration of NGC
2126 in the other 11 bands based on the method called model
calibration as presented by Zhou et al.~(\cite{zhou99}).

\section{SEARCHING FOR VARIABLES}

In the frames, about 4416 stars are detected in the field of NGC
2126. To search for variable stars, we count on magnitude scatter of
each measurement. The magnitude scatter of all stars in the {\it
i}-band is shown in Figure 1. The abscissa is mean magnitude of each
star. The dense parts are normal stars which are intrinsically
stable in Figure 1. A star is selected as a variable candidate if 1)
it presents a large deviation ($>3$ rms) compared with that of the
other stars of similar magnitude; 2) it has been measured on at
least 100 frames. Based on these conditions, hundreds of possible
variable candidates were selected. Then, we checked the real-time
light curves and the suspect time-series data of all candidates.
The candidates showing noisy chaotic light variations, were rejected
from the variable sample. After that, the number of the variable
candidates decreased to about fifties. Furthermore, we checked the
frames, and got rid of the stars that locate in the image edge, or
saturated or are contaminated by the bright stars. Ultimately, a
total of 19 certain variable stars were identified from the
candidates. Among these 19 stars, 5 variables named V1-V2,V4-V6 are
known variables from G\'asp\'ar et al.~(\cite{a.g03}), the remaining
14 objects named V7-V20 being newly discovered. A known variable
star named V3 in G\'asp\'ar et al.~(\cite{a.g03}) was not detected
as a variable candidate in this paper. We then check the data, and
found that its variable amplitudes were too low to be detected. In
addition, there are two candidates, the properties of which cannot
be determined in this paper. The positions of all 21 variable
candidiates in the field of NGC 2126 were marked in Fig. 2, and the
newly found variables are marked in sequence of the distance away
from the center of the cluster.

The analysis of the light-curves was made with the phase dispersion
minimization (PDM) method (Stellingwerf~\cite{stel78}). In this
method, the main characteristics of light curves such as the periods
and amplitudes of light variations, were given.

According to the periods, spectral types and the behaviors of the
light curves, the variable stars are classified: seven binaries
systems, five pulsating variable stars, a RS CVn star, a W UMa star
or $\delta$ Scuti star, one eclipsing binaries with pulsating
component, three long period variables, and three objects that
cannot be confirmed based on the PDM method. The analyzed results,
including their periods, variable amplitudes, mean magnitudes and
spectral types, are also listed in Table 1. The real-time light
curves of 6 variables are shown in Figure 3. The phased light curves
of V1, V2, V4, V5 and V6 are presented in Figure 4. The phased light
curves of 11 new variables are shown in Figure 5. The phased light
curves of V22 are shown in Figure 6.

\section{The Physical Parameters and CMD of The Cluster}

Based on the photometric data in the BATC 13 filter bands, the
fundamental parameters of the open cluster NGC 2126 are determined
by fitting the spectral energy distributions (SEDs) of stars in the
cluster field with theoretical SEDs of Padova stellar evolutionary
models (Girardi et al.~\cite{gir00},~\cite{gir02}; Wu et al.~\cite{wu05}). 
At the same time, the membership probabilities of the stars including the variable stars studied in
this paper, were determined based on the method of Wu et
al.~(\cite{wu06}). The best-fitting theoretical results of the
cluster NGC 2126 are: the metallicity $Z$=0.008$Z_\odot$ , the age
$\log(t)$=8.95, the distance modulus $(m-M)_{0}$=10.34
and the reddening value $E(B-V)=0.55$ mag. {\bf The derived reddening is much large than that derived by G\'asp\'ar et al.~(\cite{a.g03})}

{\bf Malkov \& Kilpio~(\cite{mal02}) used various observational data to common interstellar extinction models. They recommended to use the model proposed by Arennou et al.~(\cite{aren92}) to calculate the extinction for object with distance great than 1 kpc and with galactic latitude $|d|$= 30$^\circ$. Adopting the distance 1.170 kpc of NGC 2126 calculated from above derived $(m-M)_{0}$=10.34, the visual extincation $A_v$= 1.467$\pm$0.369 is determined based on the model of Arennou et al.~(\cite{aren92}). Adopting $R_v$= 3.3, the extioncation of NGC 2126 is estimated to be $E(B-V)$= 0.45$\pm$0.11, which is consistent with that derived in this study within the errors.}


Also based on the photometric data in the BATC 13 filter bands, the
spectral types of the stars including the variable stars studied in
this paper, can be given by fitting the BATC SEDs with the
module SEDs of Gunn \& Stryker~(\cite{gunn83}), in which 74 spectral
types from O to M6 and luminosities from main sequence to giant are
included. 

The results of membership probabilities are shown in the last column in
 Table 1. In this table, "-" denotes that we did not obtain this parameter of
this variable in this paper. The CMD of the observed field is shown
in Figure 7. The filled circles denote the stars whose membership
probabilities are more than 70 percent. The black squares denote the
variable stars. In addition, because V10, V14, V16 and V19 do not
lie $e$-band in the image, they are not plotted in the CMD
diagram. The thick solid line denotes the theoretical isochrones
with $\log(t)$ = 8.95. The theoretical isochrones with the parameters derived
 by SEDs-fitting can fit star distributions in the CMD very well.
The membership probabilities of V1 and V2 are 0.91 and 0.65,
respectively. They are on the main sequence of the cluster NGC 2126.
Therefore, V1 and V2 should be the members of NGC 2126.
 Considering the characteristics of variable stars and the positions
in the CMD (Fig. 7), V5 and V13 may be members of NGC 2126, although
its membership probabilities are only 0.30 and 0.46, respectively. V7 locates
near the turn-off, and should be a member of NGC 2126.

\section{RESULTS AND DISCUSSIONS}

\subsection{The known variables in NGC 2126}

There are a total of 5 known variable stars (G\'asp\'ar et al.~\cite{a.g03})in 
the field of NGC 2126. Their basic parameters are listed in Table 1. During 
this observation, the good light curves for V1, V2, V4 ,V5 and V6 were
recorded.

V1: Being similar to the previous work, we obtained two periods of
V1, 0.82222 d and 1.64444 d comparing with 0.822235 d and 1.6447 d
from G\'asp\'ar~(\cite{a.g03}). G\'asp\'ar~(\cite{a.g03}) suggested
that V1 could be $\gamma$ Dor or RS CVn star. The double-peak of the
phase diagram is clear with the period of 1.64444 d (Fig. 4). In our
observation, we classified V1 as A-type. However, the RS CVn
criteria, which was defined by Hall~\cite{ha76} and is widely used,
suggests that the spectral type of RS CVn star is F$\sim$K type.
Kaye et al.~(\cite{kaye99}) defined the $\gamma$ Dor phenomenon that
consists of variable stars with an implied range in spectral type
A7-F5 and in luminosity class IV, IV-V or V. Therefore, V1 may be
$\gamma$ Dor star rather than a RS CVn.

V2: It was discovered by G\'asp\'ar~(\cite{a.g03}), and the periods
around 0.5 d and 1 d were given, however neither period cannot give
a continuous phase diagram with no gap. So,
G\'asp\'ar~(\cite{a.g03}) did not determine the cause of the
low-amplitude. In this paper, its spectral type is decided to be
A-type. The phased light curves of Figure 4 show that V2 is becoming
faint and its period may be longer than 2 day.

V4: It is an Algol-type eclipsing binary. G\'asp\'ar~(\cite{a.g03})
observed only one minimum of this star and failed to determine its
period. During our observations, we recorded three eclipse events.
Using the PDM code, a probable period of 1.0966 d is derived. In
addition, we determined its spectral type to be F-type.

V5: G\'asp\'ar~(\cite{a.g03}) analyzed the frequencies of this star
with Fourier analysis, and determined two frequencies $f_{1}$
=$11.43
~d^{-1}$ and $f_{2}$=12.14 $d^{-1}$ and $f_{1}$/$f_{2}=0.94$. So
they suggested V5 was a multiply periodic oscillations $\delta$
Scuti star with non-radial modes. In this study, its spectral type
is classified as A-type. Based on the PDM code, we derive a probable
period to be 0.0827 d which is close to the $f_{1}$, and the
amplitude is 0.104 mag, very low. Combining the results of
G\'asp\'ar~(\cite{a.g03}) and of this paper, V5 may be confirmed to
be $\delta$ Scuti star.

V6: G\'asp\'ar~(\cite{a.g03}) showed that V6 was an eclipsing binary
system with a $\delta$ Sct-type pulsating with $P_{pul}=0.129$ d and
$P_{orb}=1.1732$ d, and it might have $f_{pul}:f_{orb}=1:9$
resonance between the orbital motion and pulsation. Our results of
$P_{pul}=0.12957$ d and $P_{orb}=1.1732$ d are consistent with
G\'asp\'ar~(\cite{a.g03}). Considering its spectral type of F-type
determined in this paper, V6 may be a $\delta$ Scuti star in an
eclipsing binary.

\subsection{The newly discovered variables}

In this paper, sixteen new variables were discovered in the field of
NGC 2126. According to the behaviors of the light curves, the
spectral types and the period analysis, classifications were made.
Among the sixteen new variables, four were classified as eclipsing
systems, one as W UMa type star, one as EA star, one as W UMa star
or $\delta$ Scuti star, three as pulsating variable stars (one RR
Lyr and two $\delta$ Scuti), one RS CVn star, three long period
variables, and two variable candidates that are not confirmed in
this paper. Their basic parameters are listed in Table 1. The
real-time light curves of V11-V13, V15-V16 and V19 are shown in Fig.
3. The phased light curves of V7-V11, V14, V17-V21 are shown in Fig.
5. The phases light curves of V22 are plotted in Fig. 6.

V7: Its position in the CMD , i.e. it is near the turn-off, shows,
that it is a member of this cluster. It is difficult to determine
its variable type. From the period of 0.395 d and the behaviors of
the light curves, it looks like a W UMa binary system, which has a
type of G or K and periods of 0.22 to 0.8 days
(Binnendijk~\cite{bin65}). However, according to the spectral type
of A-type, and the light curves of sinusoidal variations, V7 may be
a high amplitude $\delta$ Scuti star. The spectral types of $\delta$
Scuti stars are from A to F with short periods of 0.02 to 0.3 days
(Breger~\cite{bre91}).

V8: It is likely an eclipsing binary system with period of 0.28448
d. The phased light curves are plotted in Figure 5.

V9: It posses typical W UMa-type light curves, and its spectral type
is classified as a F-type, which consists with W UMa type, i.e.
F$\sim$G. Therefore, V9 should be a W UMa eclipsing system with a
period 0.444462 d.

V10: This star may be a field eclipsing system, the period of which
is 0.47743 d. The phased light curves are shown in Figure 5.

V11, V13 and V15: They may be field long period variable stars. As
we know, the spectral types of red giants and supergiants are M, R, N or S, and the periods are up to 600 days. In our observations,
V11 ,V13 and V15 are all classified as M-types. Their real-time
light curves also have long time variations (Fig. 3). Based on PDM
code analysis, we obtained a probable period of V11 to be 38.5 d.
The phase diagrams are plotted in Figure 5. The periods of V13 and
V15 cannot be obtained in this paper.

V14: It is a field Algol-type eclipsing system. Its light curves
have deep primary minima and flat bright parts. A period of 0.80042
d is obtained.

V17: It is classified as A-type in this paper. The light curves
suggest it is likely a $\delta$ Scuti variable with a period 0.1272
d.

V18: About V18, its spectral type is classified as a K-type in this
study. Analysis of the light curves reveals that it is likely an
eclipsing binaries system with a period 0.20042 d.

V19: Its real-time light curve posses the character of RR Lyra,
asymmetrical, and increasing rapidly and decreasing slowly. So, V19
may be a RR Lyra system with $P=0.5375$ d.

V20: It is classified as an F - type star in this paper. From its
light curves (Fig. 5), it may be a $\delta$ Scuti star with
$P=0.30315$ d.

V21: The spectral type of V21 is classified as G-type in this paper.
From \S5.1, we know that a RS CVn is a later G $\sim$ K type. So, V21
may be a RS CVn variable with a period 0.956 d. In {\it i} band, the
light curves show that it is likely a RS CVn. However, in {\it f}
band, they are much distorted.

V22: The most confused object in this paper is V22. The analysis of
light curves gives two periods, one is 0.22356 d and the other is
0.447124 d (Fig. 6). From the phased diagram of the period 0.22356
d, V22 is most likely an eclipsing contact binary system. On the
other hand, the light curves of the period 0.447124 d have a deep
primary minima and a second minima. So, V22 may be field eclipsing
binary systems.



\section{CONCLUSIONS}

In this paper, we presented a time-series CCD photometric survey of
variable stars in the field of the open cluster NGC 2126. In this
survey, 21 variables were discovered, including 5 previous ones
(G\'asp\'ar et al.~\cite{a.g03}). Based on the SED fitting between
the theoretical isochrones and the photometry, the membership
probabilities of stars and the fundamental parameters of
NGC 2126 were determined. The important results of this study are
the following:

1. We discovered 16 new variable stars in this survey: six were
classified as eclipsing systems with four eclipsing binary systems
(V8, V10, V18 and V22), a W UMa type star (V9), an EA star (V14),
three long period variables (V11, V13 and V15), three pulsating
variable stars with a RR Lyr (V19) and two $\delta$ Scuti (V17 and
V20), and a RS CVn star (V21). About V7, its light curves show that
it is most likely a W UMa type star. However, its spectral type is
classified as A-type. So, it is very difficulty to confirm whether
it is a W UMa star or $\delta$ Scuti star. The more information will
be needed in the future observation. In addition, two variable
candidates(V12 and V16) were not confirmed in this paper. Among the
16 newly discovered variable stars, two (V7 and V13) may be the
members of the open cluster NGC 2126.

2. We also determined the mass of each variable star belonging to
members of cluster NGC 2126, and the physical parameters of this
cluster. The masses of V1, V2, V5 and V13 are 1.1312, 0.9131, 1.4808
and 0.4754$M_\odot$, respectively. The mass of V1 suggests that it
may locate at the blue edge of the instability strip; and its
pulsations are driven by the modulation of the radiative flux from
the interior of the star that is a relatively deep envelop
convection zone (Guzik et al.~\cite{guzi00}; Warner et
al.~\cite{war03}). As a $\gamma$ Dor, the variations of V1 are
consistent with the model of high order ($n$) and low spherical
degree ($l$), nonradial, gravity ($g$)-mode oscillations (Kaye et
al.~\cite{kaye99}). Because we cannot determine the pulsating class
of V2, its pulsating reason is unknown. Although the spectral type
of V13 is known, the real-light curves are only a part of the whole,
and the driving mechanism is not known.

3. The physical parameters of the open cluster NGC 2126 are: the
metallicity $Z=0.008Z_\odot$, the age $\log(t)=8.95$, the
distance modulus $(m-M)_{0}=10.34$ and the $E(B-V)=0.55$ mag.

\begin{acknowledgements}
This work has been supported by the Chinese National Natural Science
Foundation through Grant Nos. 10873016, 10803007, 10473012, 10573020, 10633020, 10673012, 10603006, 10773015, 10778720; and by National Basic Research Program 
of China (973 Program) No. 2007CB815400; and the
Knowledge Innovation Program of the Chinese Academy of Sciences. We
thank the colleagues of BATC for the  assistance of observation and
useful discussion.
\end{acknowledgements}


\clearpage

\begin{figure}
\vspace{2mm}
\begin{center}
\hspace{3mm}\psfig{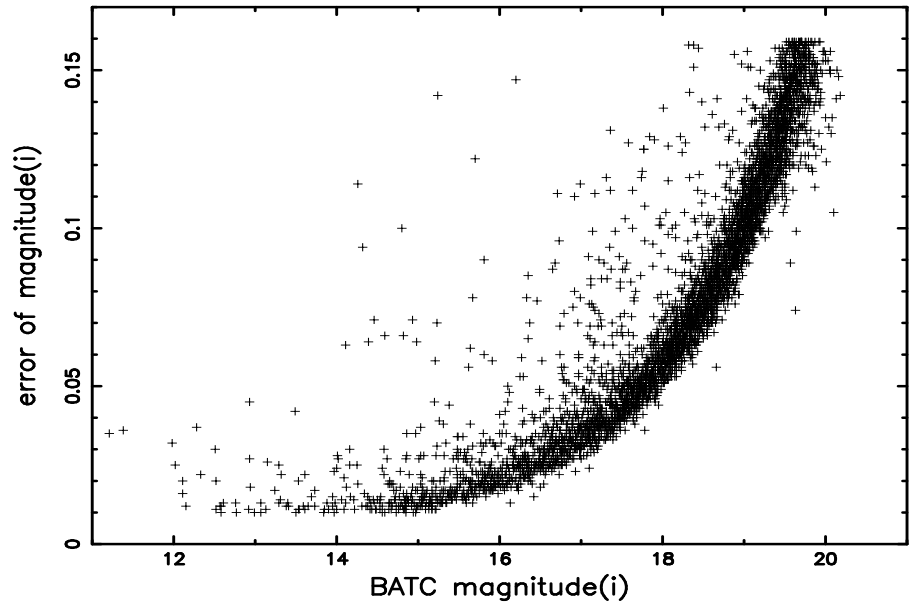}
\caption{Plotting of the rms scatter of {\it i}-band for the BATC
observation in the field of NGC 2126. The dense parts are normal
stars, which are intrinsically stable.} \label{Fig:fig1}
\end{center}
\end{figure}

\begin{figure}
\vspace{4mm}
\begin{center}
\hspace{3mm}\psfig{figure=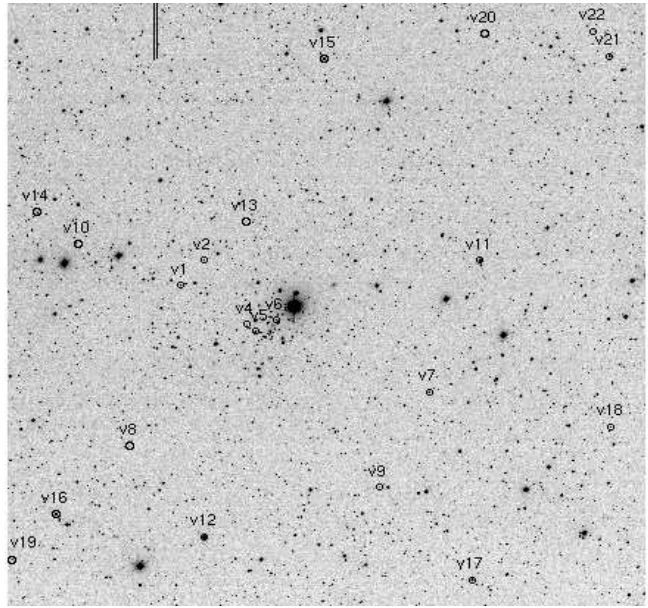,width=120mm,height=80mm,angle=0.0}
\caption{Observed CCD field ($58{\arcmin} \times 58{\arcmin}$) of
the open cluster NGC 2126. Identification of variable stars are also
marked. North is up and east is to the left} \label{Fig:fig2}
\end{center}
\end{figure}

\begin{figure}
\centering
\includegraphics[width=0.97\textwidth,height=35mm]{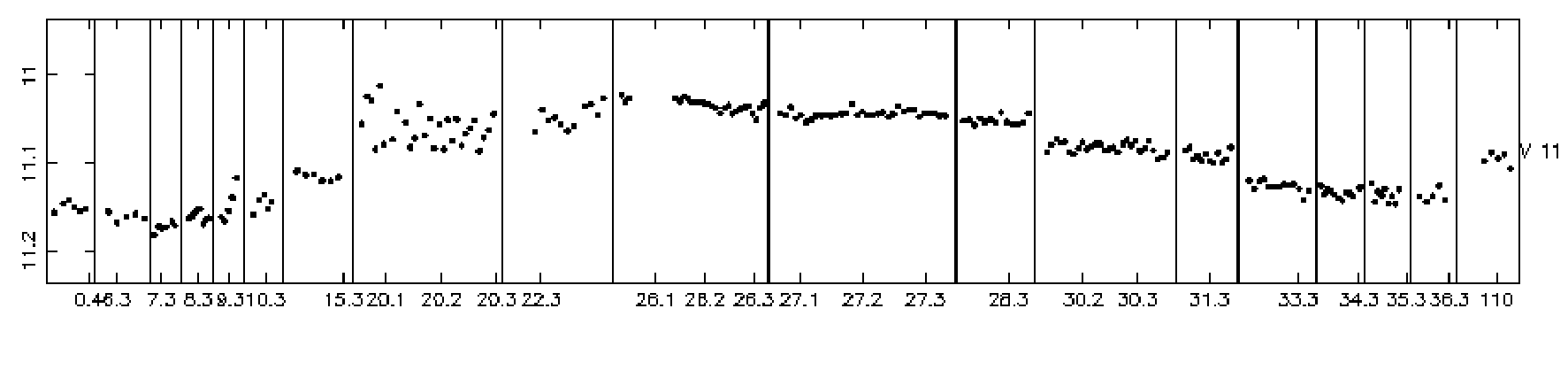}
\includegraphics[width=0.97\textwidth,height=35mm]{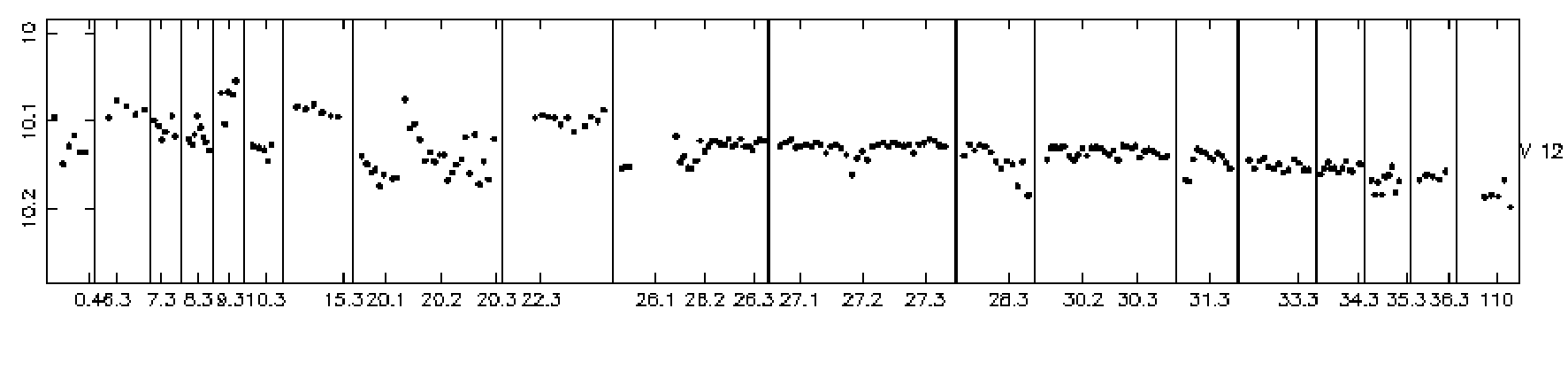}
\includegraphics[width=0.97\textwidth,height=35mm]{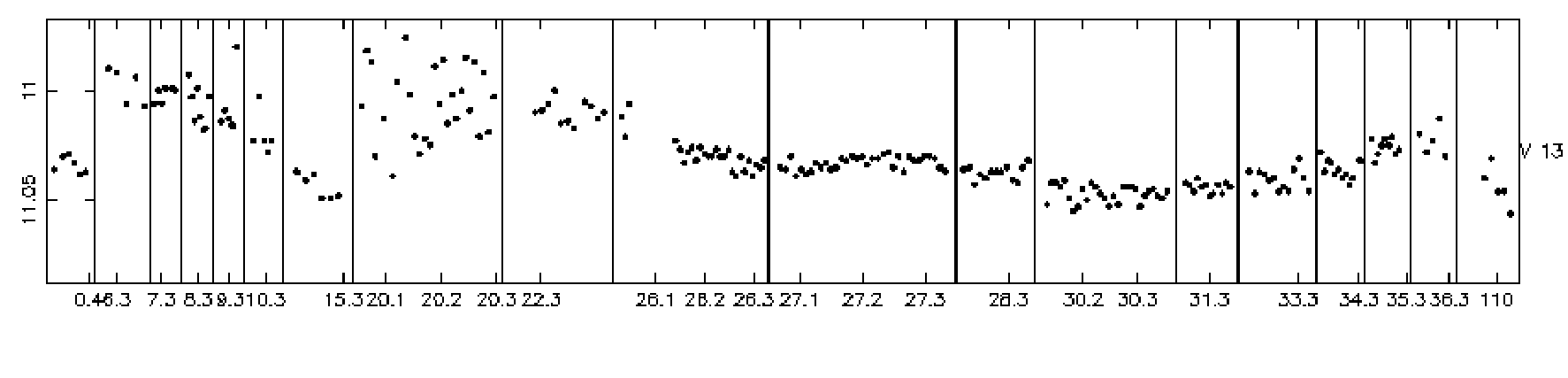}
\includegraphics[width=0.97\textwidth,height=35mm]{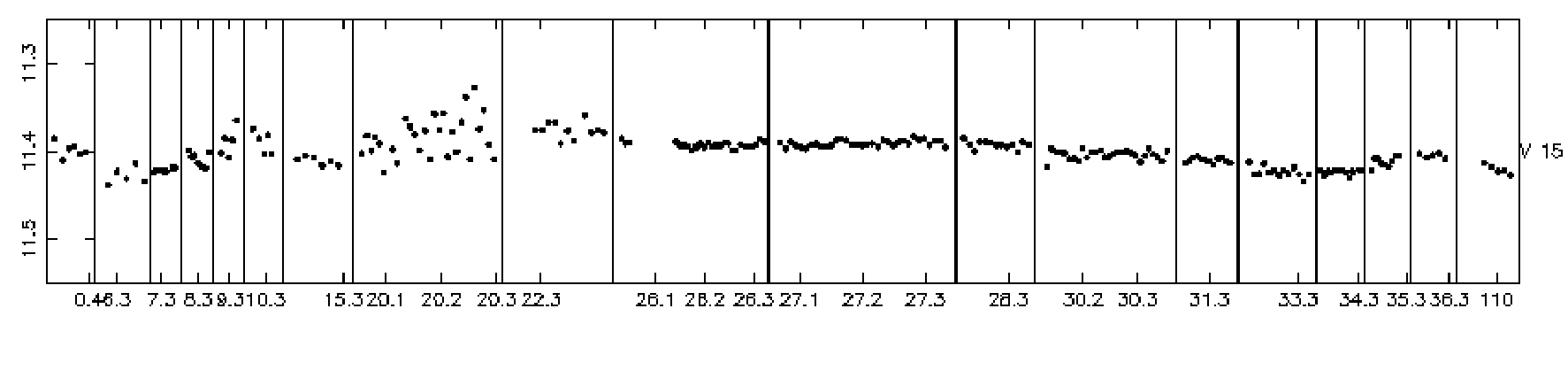}
\includegraphics[width=0.97\textwidth,height=35mm]{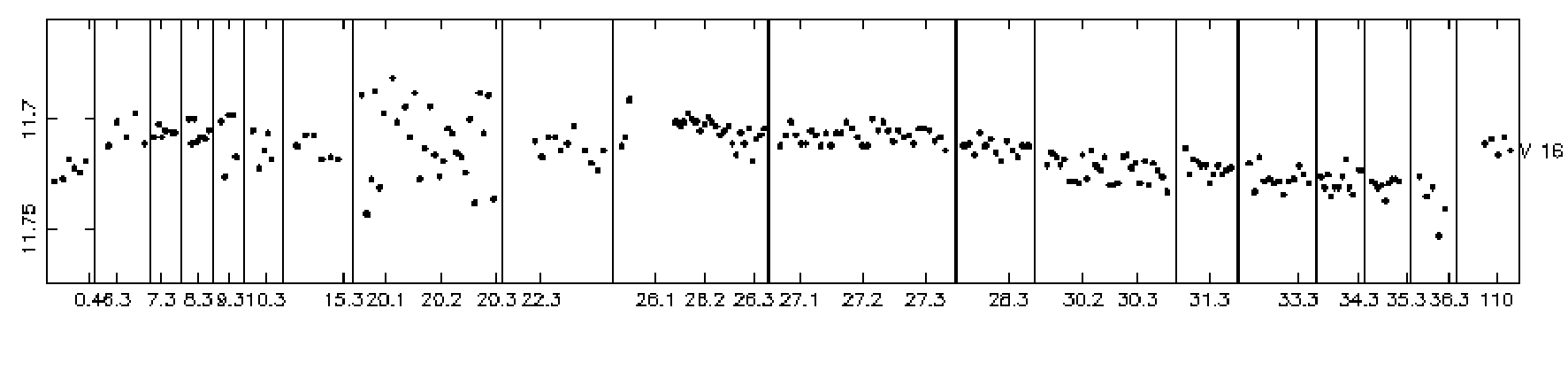}
\includegraphics[width=0.97\textwidth,height=35mm]{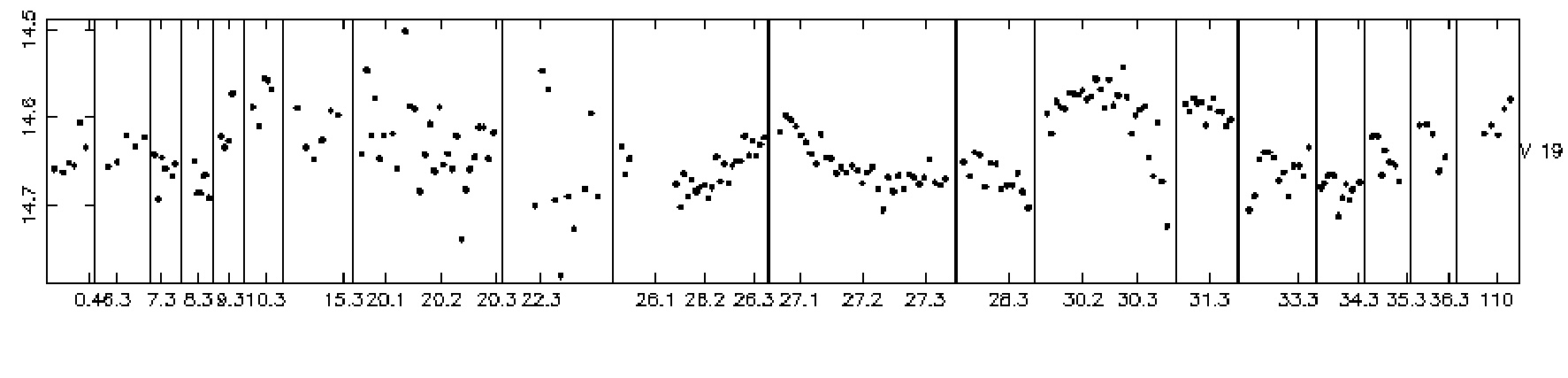}
\caption{Real-time light curves for six variable stars.}
\label{Fig:fig3}
\end{figure}

\begin{figure}
\vspace{1mm}
\hspace{10mm}
\includegraphics[angle=270,scale=.35]{v1-1.ps}
\includegraphics[angle=270,scale=.35]{v1-2.ps}
\includegraphics[angle=270,scale=.35]{v2.ps}
\end{figure}
\begin{figure}
\vspace{1mm}
\hspace{10mm}
\includegraphics[angle=270,scale=.35]{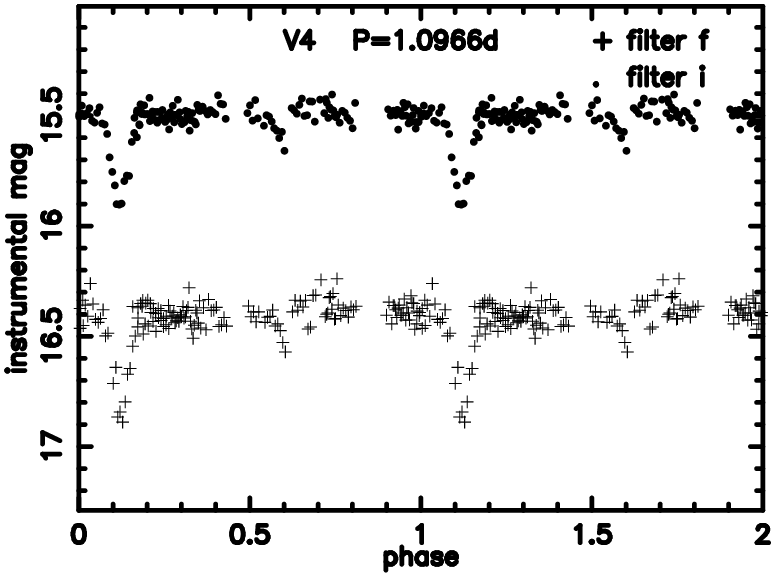}
\includegraphics[angle=270,scale=.35]{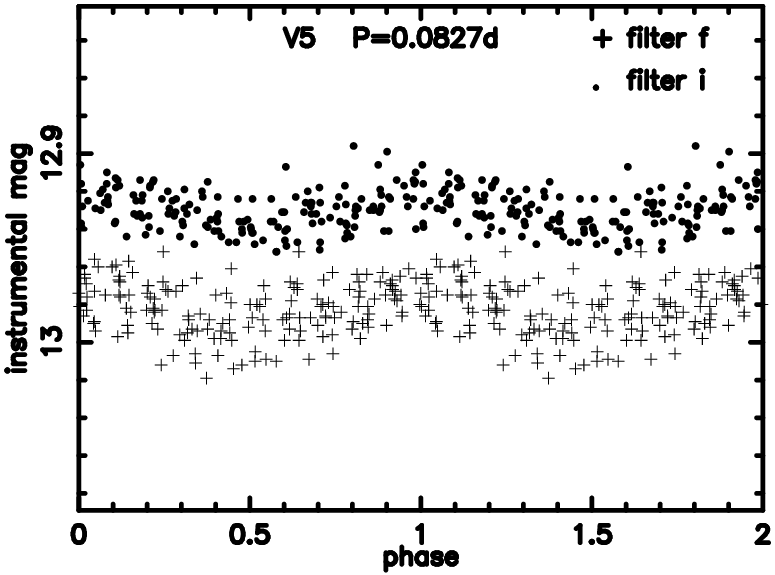}
\includegraphics[angle=270,scale=.35]{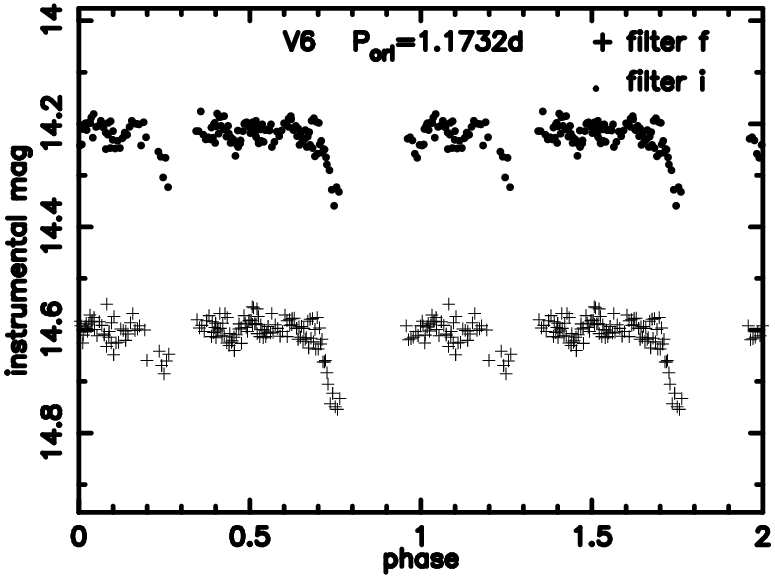}
\end{figure}
\begin{figure}
\vspace{1mm}
\hspace{10mm}
\includegraphics[angle=270,scale=.35]{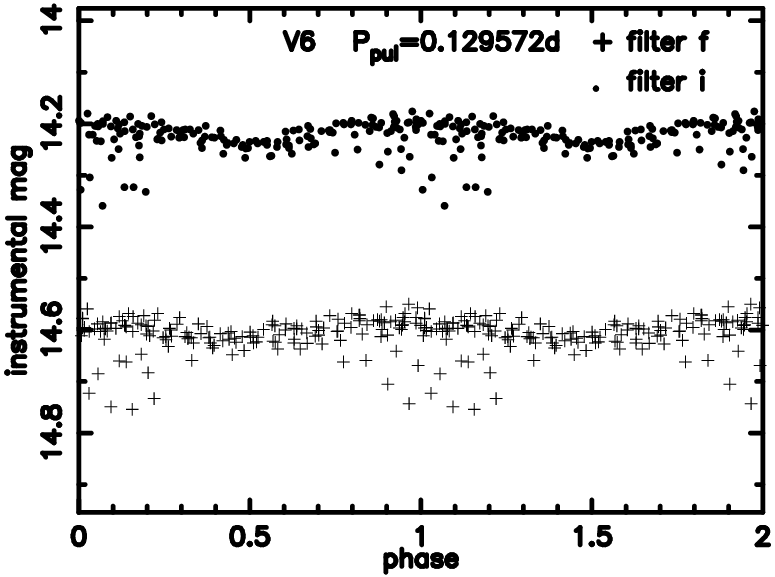}
\caption{The light curves for 5 known variable stars in {\it f} and
{\it i} bands, respectively, with the periods marked in each
frame.}\label{fig4}
\end{figure}

\clearpage

\begin{figure}
\vspace{1mm}
\hspace{10mm}
\includegraphics[angle=270,scale=.35]{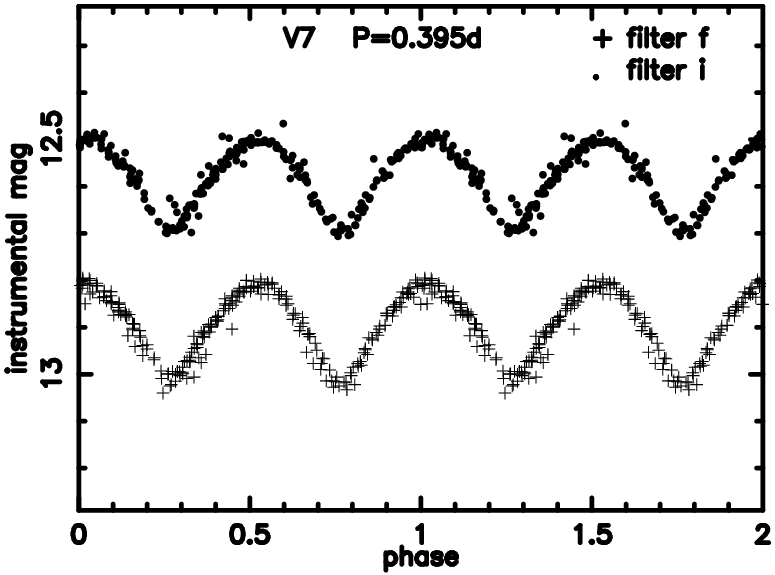}
\includegraphics[angle=270,scale=.35]{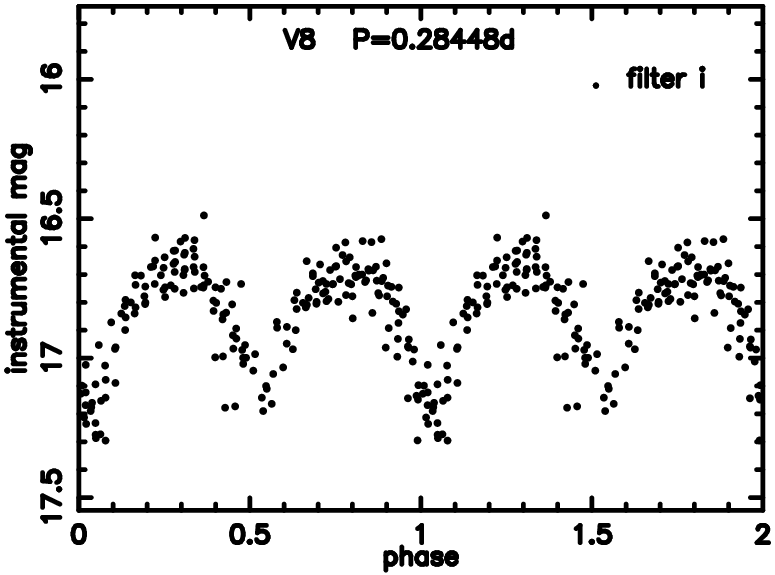}
\includegraphics[angle=270,scale=.35]{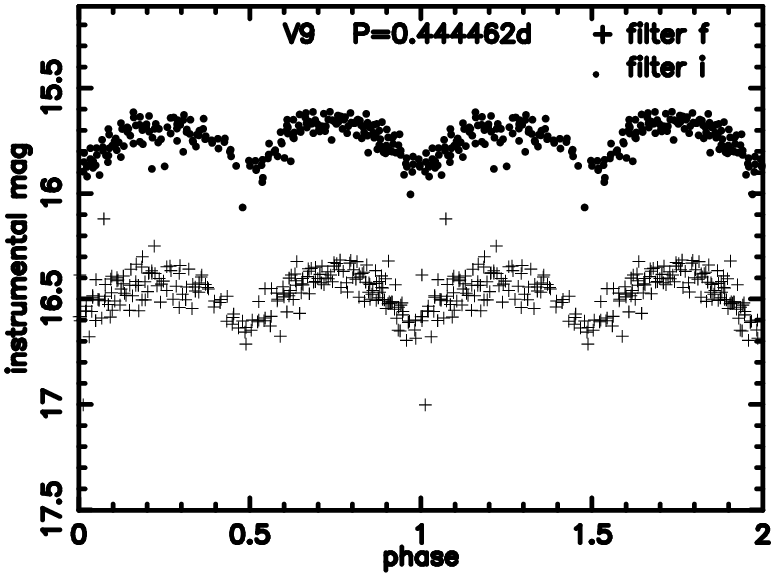}
\end{figure}
\begin{figure}
\vspace{1mm}
\hspace{10mm}
\includegraphics[angle=270,scale=.35]{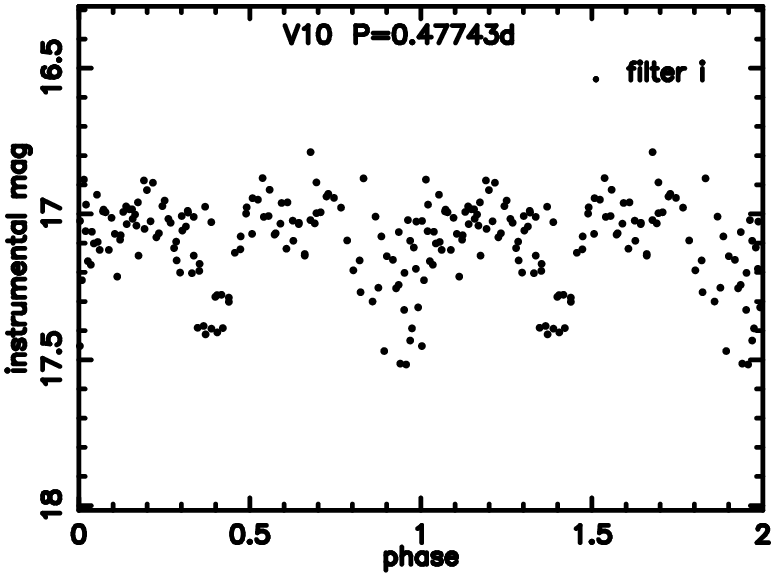}
\includegraphics[angle=270,scale=.35]{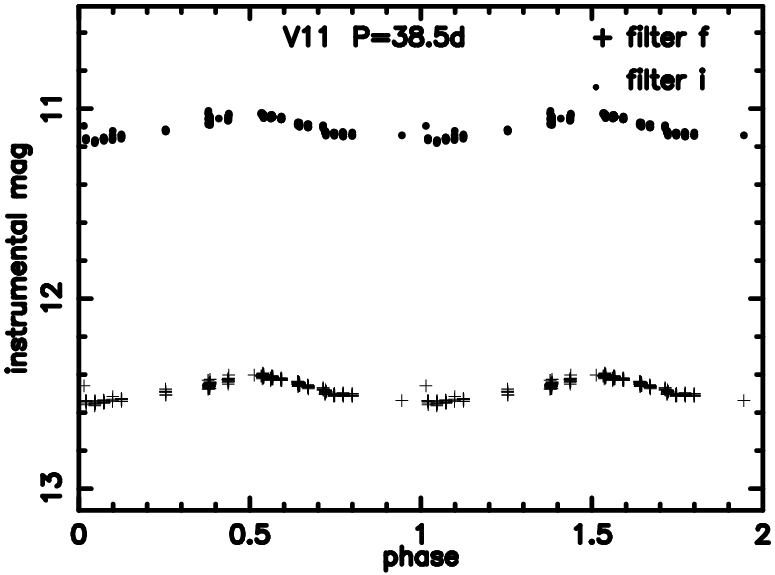}
\includegraphics[angle=270,scale=.35]{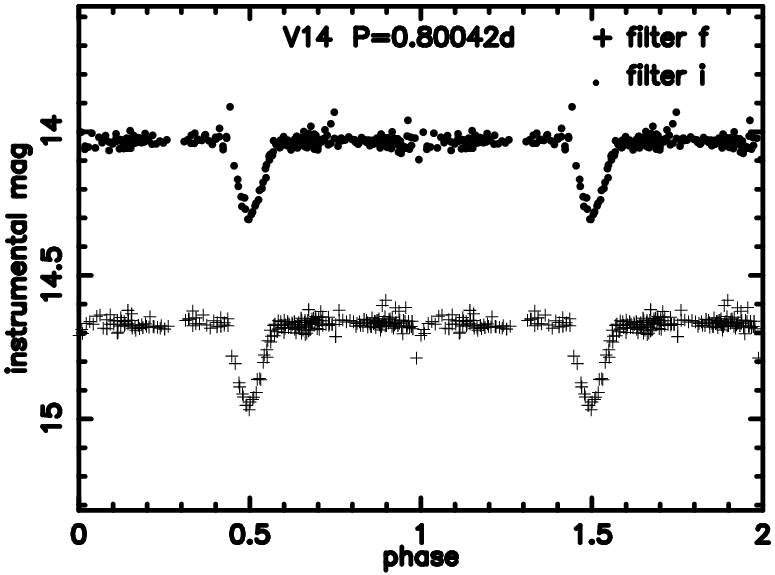}
\end{figure}
\begin{figure}
\vspace{1mm}
\hspace{10mm}
\includegraphics[angle=270,scale=.35]{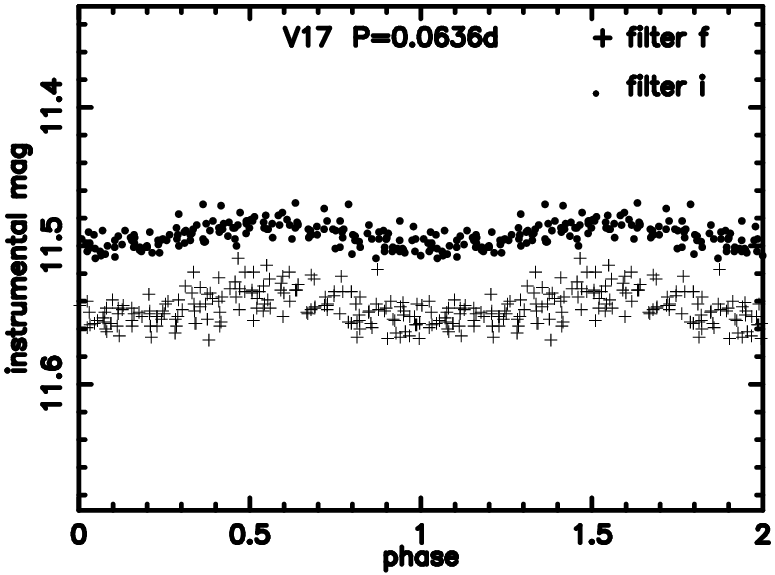}
\includegraphics[angle=270,scale=.35]{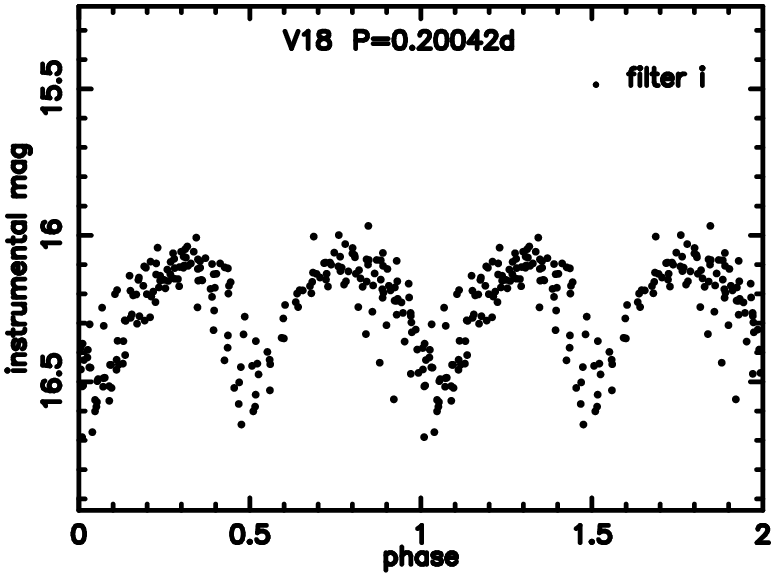}
\includegraphics[angle=270,scale=.35]{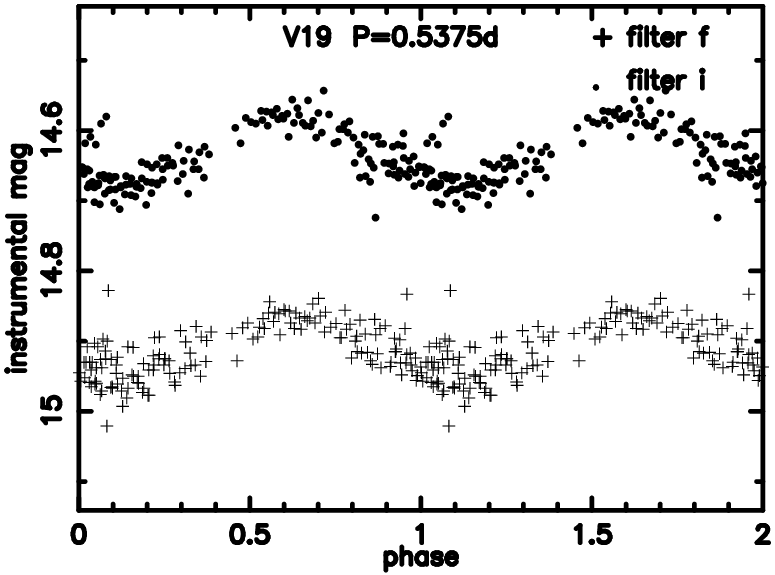}
\end{figure}
\begin{figure}
\vspace{1mm}
\hspace{10mm}
\includegraphics[angle=270,scale=.35]{v20.ps}
\includegraphics[angle=270,scale=.35]{v21.ps}
\caption{The light curves for new variable stars in {\it f} and {\it
i} bands, respectively, with the periods marked in each
frame.}
\label{Fig.5}
\end{figure}

\begin{figure}
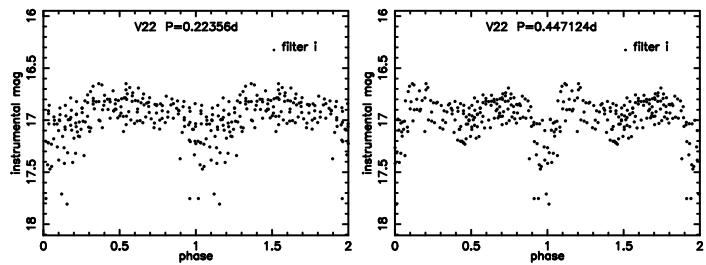

\vspace{1mm}
\hspace{10mm}
\includegraphics[angle=270,scale=.35]{v22-1.ps}
\includegraphics[angle=270,scale=.35]{v22-2.ps}
\caption{The light curves for v22 with two possible periods in {\it
i} bands.}
\label{Fig.6}
\end{figure}

\begin{figure}
\centering
\includegraphics[angle=270,scale=.90]{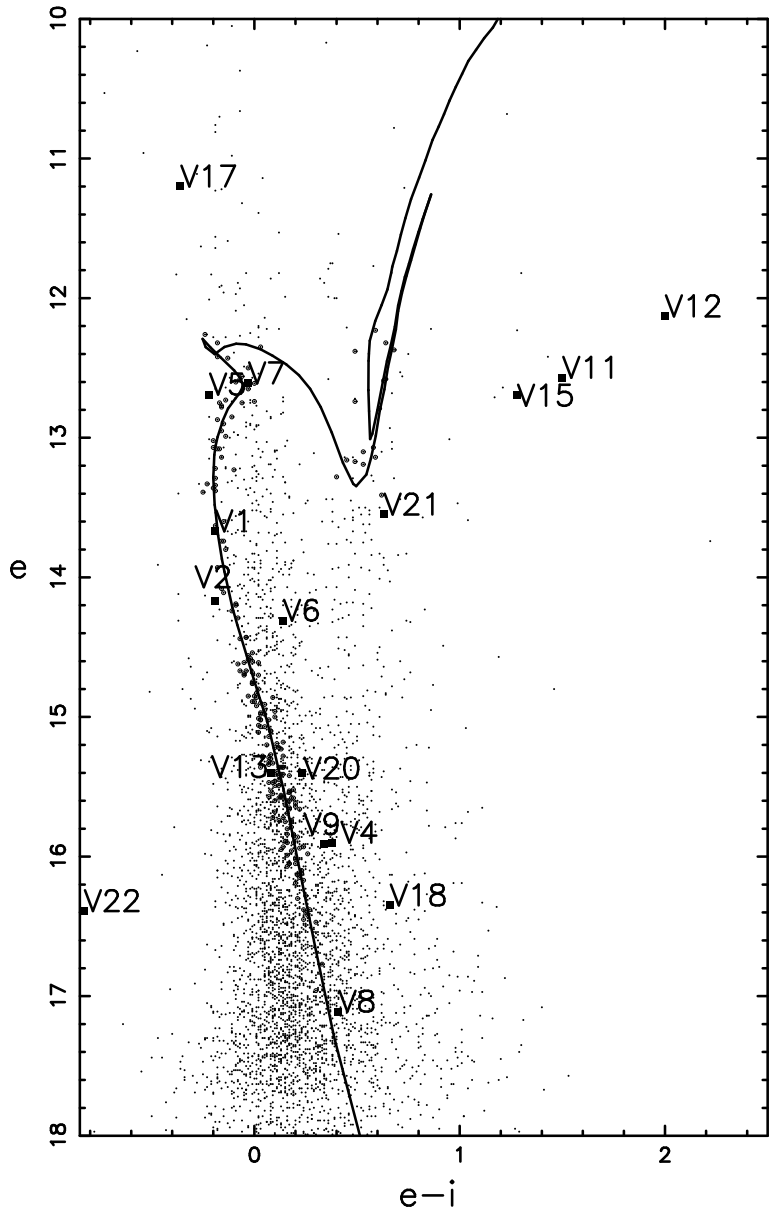}
\caption{The color-magnitude diagrams of NGC 2126. 
The solid line show the $\log(t)$=8.96 theoritical isochrone (Girardi et
al.~\cite{gir00}). The filled circles denote stars observed whose
probabilities to be members are more than 70 percent. The black squares
denote the variabel stars.}
\label{Fig.7}
\end{figure}

\begin{table}[tbp]
 \caption[]{Observation summary 21 variable star in open cluster NGC 2126.}
 \begin{center}\begin{tabular}{llllllclcl}
 \hline\noalign{\smallskip}
ID & Period  & $\alpha_{2000}$ & $\delta_{2000}$ & $\bar{m}$({\it i})
& error & $\Delta$amp({\it i}) & Spec & Variable Type & Memeber          \\
 \hline\noalign{\smallskip}
V1& 1.64444 & 06:01:44.09 & +49:56:30.3 & 13.868 & 0.02& 0.198 & A & $\gamma$ Dor & 0.91\\
  &or 0.82222 &\\
V2 &-  & 06:01:57.36 & +49:58:54.8\ & 14.356 & 0.025& 0.117 &  A & -& 0.65\\
V4 & 1.0966 & 06:02:21.28 & +49:52:37.0 & 15.507 & 0.091& 0.637 & F & EA & 0.00\\
V5 & 0.0827 & 06:02:26.38 & +49:51:56.5 & 12.930 & 0.05 & 0.104 & A & $\delta$ Sct & 0.30\\
V6 &1.1732 & 06:02:37.99 & +49:53:02.5 &  14.211& 0.045 &0.479 & F & EA+$ \delta$Sct &0.00\\
   &or 0.129572 &\\
V7 &0.395 & 06:04:04.14 & +49:45:53.4 & 12.671 & 0.066 & 0.240 & A & W UMa or $\delta$ Scuti & -\\
V8 &0.28448 & 06:01:14.33 &+49:40:58.9 &16.364  &0.184 &0.808 &- &eclipsing binaries & 0.00\\
V9 &0.444462 & 06:03:34.89 & +49:36:45.0 & 15.626 & 0.083& 0.454 & F & W UMa & 0.00\\
V10 &0.47743 & 06:00:46.04 & +50:00:34.9 & 17.032 & 0.163& 0.728 & - & eclipsing binaries & 0.00\\
V11 & 38.5 & 06:04:34.69 & +49:58:33.4 & 11.032  & 0.045& 0.170 & M & LPV & 0.00\\
V12 &- & 06:01:55.94 &+49:32:08.5 &10.122 &0.024 &0.134 &- &uncertain& 0.00\\
V13 & - & 06:02:21.61 & +50:02:35.5 & 11.007 & 0.018 & 0.090 & M & LPV & 0.46\\
V14 & 0.80042 & 06:00:22.78 & +50:03:40.4 & 14.013 & 0.058 & 0.392 & - & EA & -\\
V15 & - & 06:03:07.49 &+50:18:13.8 &11.395 &0.017 &0.111 & M  & LPV & 0.00\\
V16 & - & 06:00:32.51 & +49:34:27.4 & 11.692 & 0.012 & 0.085 & - & uncertain& -\\
V17 &0.0636 & 06:04:26.43 &+49:27:39.5 &11.494 & 0.012& 0.093 & A & $\delta$ Scuti & 0.00\\
V18 &0.2004 & 06:05:46.45 &+49:42:11.0 &16.084 &0.160 &0.721 & K & eclipsing binaries& - \\
V19 & 0.5375 & 06:00:07.54 & +49:30:07.5 & 14.640 & 0.042 & 0.279 & - & RRa Lyr & -\\
V20 & 0.30215 & 06:04:39.65 & +50:20:25.8 & 15.228 & 0.111 & 0.709 & F & $\delta$ Scuti & -\\
V21 &0.956 & 06:05:50.90 &+50:18:01.8 &12.961 &0.022 &0.169 & G & RS CVn & 0.02\\
V22 & 0.22356& 06:05:42.28 & +50:20:27.0 & 16.642 & 0.201 & 1.158 & -& eclipsing system & 0.00\\
  & or 0.447124 &\\
\noalign{\smallskip}\hline
  \end{tabular}
\end{center}
\end{table}

\label{lastpage}


\begin{thebibliography}{}
\bibitem[1992] {aren92} Arenou F.. et al., 1992, \aap,   258, 104

\bibitem[2003] {a.g03} G\'asp\'ar A. et al., 2003, \aap,   410, 879

\bibitem[1991] {bre91} Breger M., 1991, \aap, 250, 109

\bibitem[1965] {bin65} Binnendijk L., Veroeffentlichungen der
Remesis-Sternwrte au Bamberg, Nr 40;, P. 36, 1965

\bibitem[1943] {cuf43} Cuffey J., 1943,  \apj, 97, 93

\bibitem[2007] {mar07} De Marchi F., Poretti E. et al., 2007,
\aap, 471, 526

\bibitem[1996] {fan96} Fan X. H. et al., 1996, \aj, 112, 628

\bibitem[1998] {fran98} Frandse S., Arentoft T., 1998, \aap, 333, 524

\bibitem[1996]{fuk96} Fukugita M. et al., 1996, \aj, 111, 1748

\bibitem[2000]{gir00} Girardi L., Bressan A., Bertelli G., Chiosi C., 2000a, \aap, 141, 371

\bibitem[2002]{gir02} Girardi L., Bertelli G., Bressan A. et al., 2002, \aap, 391, 195

\bibitem[2000] {guzi00} Guzik J. A., Kaye A. B., Bradley P. A., Cox A. N., Neuforge C., 2000, \apj, 542, L57

\bibitem[1983]{gunn83} Gunn J. E., Stryker L. L., 1983, \apjs, 52, 121

\bibitem[1976]{ha76}  Hall D. S., 1976, in IAU Colloquium No. 29,
"Multiple Variable Stars" (D. Reudel:Boston),p.278-348

\bibitem[2005] {hu05} Juei-Hwa H., Wing-Huen et al., 2005, \cjaa, 4, 356

\bibitem[2007] {kang07} Kang Y. B., Kim S. -L., Rey S. C., 2007, \pasp, 119, 238

\bibitem[1999] {kaye99} Kaye A. B., Handler G., Krisciunas K.,  Poretti E., 1999, \pasp, 111, 840



\bibitem[2004] {li04} Li L. et al., 2004, \cjaa, 5, 411

\bibitem[1987]{lyn87} Lyng{\aa} G., 1987, Fifth catalogue of cluster parameters, Strasbourg

\bibitem[1995] {maz95} Mazur B. et al., 1995, \mnras, 273, 59

\bibitem[2002] {mal02} Malkov O., Kilpio E., 2002, \apss, 280,115

\bibitem[1983]{oke83} Oke J. B.,  Gunn J. E., 1983, \apj, 266, 713

\bibitem[2000]{par00} Park N., Nemec J. M., 2000, \aj, 119, 1803

\bibitem[1978] {stel78} Stellingwerf R. F., 1978, \apj, 224, 953

\bibitem[1987] {stet87} Stetson P. B., 1987, \pasp, 99, 191


\bibitem[2003] {war03} Warner P. B., Kaye A. B., Guzik J. A., 2003, \apj, 593, 1049

\bibitem[2002] {wu02} Wu Z. Y. et al., 2002, \aap, 381, 464

\bibitem[2005] {wu05} Wu Z. Y., zhou X., Ma J. et al., 2005, \pasp, 117, 32

\bibitem[2006] {wu06} Wu Z. Y. et al., 2006, \pasp, 118, 1104

\bibitem[2007]{wu07} Wu Z. Y. et al., 2007, \aj, 133, 2061


\bibitem[2007] {zlo07} Zloczewski K., Kaluzny J. et al., 2007, \mnras, 80, 1191

\bibitem[2004] {zhang04} Zhang X. B. et al., 2004, \mnras, 355, 1369

\bibitem[2002] {zhang02} Zhang X. B. et al., 2002, \aj, 123, 1548

\bibitem[1999] {zhou99} Zhou X., Chen J. S. et al., 1999, \pasp, 111, 909


\bibitem[2001] {zhou01} Zhou X., Jiang Z. J. et al., 2001, \cjaa, 1, 372


\bibitem[2004] {zhou04} Zhou X., Burstein D. et al., 2004, \aj, 127, 3642

\end{thebibliography}
\end{document}